\documentclass[a4paper,fleqn,usenatbib]{mnras}

\usepackage[T1]{fontenc}
\usepackage{ae,aecompl}

\usepackage{graphicx}	
\usepackage{amsmath}	
\usepackage{amssymb}	

 \newcommand{\be}{\begin{equation}}
 \newcommand{\ee}{\end{equation}}
 \def\la{\buildrel < \over {_{\sim}}}
 \def\ga{\buildrel > \over {_{\sim}}}
 
 \def\fermilat{{\it Fermi}-LAT }

\title[On the radial distribution of Galactic CR]{On the radial distribution of Galactic cosmic rays}

\author[S. Recchia, P. Blasi and G. Morlino]{
	S. Recchia,$^1$\thanks{E-mail: sarah.recchia@gssi.infn.it}		
	P. Blasi,$^{2,1}$\thanks{E-mail: blasi@arcetri.astro.it}
	G. Morlino,$^{1,2}$\thanks{E-mail: giovanni.morlino@gssi.infn.it}
\\
	{$^1$ \it INFN -- Gran Sasso Science Institute, viale F. Crispi 7, 67100 L'Aquila, Italy.}\\
	{$^2$ \it INAF -- Osservatorio Astrofisico di Arcetri, L.go E. Fermi, 5, 50125 Firenze, Italy.}
}

\date{Accepted XXX. Received 2016 April 26; in original form 2016 April 26}

\pubyear{2015}

\begin{document}
\label{firstpage}
\pagerange{\pageref{firstpage}--\pageref{lastpage}}
\maketitle

\begin{abstract} 
The spectrum and morphology of the diffuse Galactic $\gamma$-ray emission carries valuable information on cosmic ray (CR) propagation. Recent results obtained by analizing \fermilat data accumulated over seven years of observation show a substantial variation of the CR spectrum as a function of the distance from the Galactic Center. The spatial distribution of the CR density in the outer Galaxy appears to be weakly dependent upon the galactocentric distance, as found in previous studies as well, while the density in the central region of the Galaxy was found to exceed the value measured in the outer Galaxy. At the same time, \fermilat data suggest a gradual spectral softening while moving outward from the center of the Galaxy to its outskirts. These findings represent a challenge for standard calculations of CR propagation based on assuming a uniform diffusion coefficient within the Galactic volume.
Here we present a model of non-linear CR propagation in which transport is due to particle scattering and advection off self-generated turbulence. We find that for a realistic distribution of CR sources following the spatial distribution of supernova remnants and the space dependence of the magnetic field on galactocentric distance, both the spatial profile of CR density and the spectral softening can easily be accounted for.
\end{abstract}



\section{Introduction}
  \label{sec:intro}
Among the numerous open questions in CR physics there is an old one, known as ``the radial gradient problem'', concerning the dependence of cosmic ray intensity on Galactocentric distance: the CR density as measured from $\gamma$-ray emission in the Galactic disc is much more weakly dependent upon galactocentric distance than the spatial distribution of the alleged CR sources, modelled following pulsar and supernova remnant (SNR) catalogues. This result was obtained for the first time from the analysis of the $\gamma$-ray emissivity in the Galactic disk derived from the SAS-2 data by \cite{Stecker77} and later from the analysis of the COS-B data by \cite{Bhat86} and \cite{Bloemen86}, then confirmed by work based on EGRET data \citep{Hunter97,Strong96}.
In more recent years the existence of a ``gradient problem'' in the external region of our Galaxy has also been confirmed by data collected by \fermilat \citep{Ackermann11,Ackermann12}. 
The results of a more detailed analysis of \fermilat data, including also the inner part of the Galaxy, was published in two independent papers \citep{Acero16,Yang16} based on data accumulated over seven years. This study highlighted a more complex situation:  while in the outer Galaxy the density gradient problem has been confirmed once more, the density of CR protons in the inner Galaxy turns out to be appreciably higher than in the outer regions of the disk. Moreover this analysis suggests a softening of the CR spectrum with Galactocentric distance, with a slope ranging from 2.6, at a distance of $\sim 3$ kpc, to 2.9 in the external regions. These are very interesting results, but it is worth keeping in mind that the spatial distribution of CRs in the Galaxy is not the result of a direct measurement, in that it can only be inferred from the observed gamma ray emission through a careful modelling of the spatial distribution of the gas that acts as target for nuclear collisions and by separating the gamma ray contribution due to hadrons from that due to leptonic interactions (bremsstrahalung and inverse Compton scattering).

The scenario that emerges from this analysis is difficult to reconcile with the standard approach to CR propagation, which is based upon solving the transport equation under the assumption that the diffusive properties are the same in the whole propagation volume \cite[see, e.g.,][]{Berezinskii90}. Within the context of this approach, several proposals have been put forward to explain the radial gradient problem. Among them:
{\it a)} assuming a larger halo size or 
{\it b)} a flatter distribution of sources in the outer Galaxy \citep{Ackermann11};
{\it c)} accounting for advection effects due to the presence of a Galactic wind  \citep{Bloemen93};
{\it d)} assuming a sharp rise of the CO-to-H$_2$ ratio in the external Galaxy \citep{Strong04};
{\it e)} speculating on a possible radial dependence of the injected spectrum \citep{Erlykin16}.
None of these ideas, taken individually, can simultaneously account for both the spatial gradient and the spectral behavior of CR protons. Moreover, many of them have issues in accounting for other observables \cite[see, e.g., the discussion in][]{Evoli12}.

A different class of solutions invoke the breakdown of the hypothesis of a spatially constant diffusion coefficient. For instance, \cite{Evoli12} proposed a correlation between the diffusion coefficient parallel to the Galactic plane and the source density in order to account for both the CR density gradient and the small observed anisotropy of CR arrival directions. \cite{Gaggero15a, Gaggero15b} followed the same lines of thought and showed that a phenomenological scenario where the transport properties (both diffusion and convection) are position-dependent can account for the observed gradient in the CR density. It is however unsatisfactory that these approaches do not provide a convincing physical motivation for the assumed space properties of the transport parameters.

In the present paper we explore the possibility that diffusion and advection in self-generated waves produced by CR-streaming could play a major role in determining the CR radial density and spectrum. The effects of self-generated diffusion has been shown to provide a viable explanation to the hardening of CR proton and helium spectra observed by PAMELA \citep{Adriani2011} and AMS-02 \citep{Aguilar15}, supporting the idea that below $\sim 100$ GeV, particle transport at the Sun's location may be dominated by self-generated turbulence \citep{Aloisio15,2012PhRvL.109f1101B}. We suggest that this could be the case everywhere in the Galaxy and explore the implications of this scenario. In the assumption that the sources of Galactic CRs trace the spatial distribution of SNRs and that the magnetic field drops at large galactocentric distances, the density of CRs and their spectrum are well described if CRs are allowed to diffuse and advect in self-produced waves. 

The paper is structured as follows. In \S~\ref{sec:CR} we show the solution of the CR transport in a 1D slab model with constant advection and with purely self-generated diffusion. In \S~\ref{sec:results} we discuss the distribution of sources and the behavior of the magnetic field in the Galactic plane and we compare our results for the CR proton spectrum with the data obtained by \fermilat. Finally we summarize in \S~\ref{sec:conc}.

\section{CR transport in self-generated turbulence} 
  \label{sec:CR}

For simplicity, we assume that for any given galactocentric distance $R$, diffusion can be described as one-dimensional, so that particles diffuse and are advected only along the $z$ direction. The transport equation for CR protons can then be written as follows:
\begin{equation} \label{eq:transport}
  - \frac{\partial}{\partial z} \left[ D(z,p) \frac{\partial f}{\partial z} \right]
  + w \frac{\partial f}{\partial z} - \frac{p}{3} \frac{\partial w}{\partial z} \frac{\partial f}{\partial p}
  = Q_0(p) \delta(z) \,,
\end{equation}
where the advection velocity is
\begin{equation}
  w(z) = {\rm sign}(z) \, v_A   \,,
\end{equation}
directed away from the disk, and the Alfv\'en speed is $v_A= B_0/\sqrt{4\pi m_p n_i}$. We assume that the ion density $n_i$ is constant everywhere in the halo, which is assumed to extend out to $|z|=H$, while the magnetic field strength is constant along $z$ but can be a function of the Galactocentric distance, $R$. 

The injection of particles occurs only in the Galactic disk (at $z=0$) and is a power law in momentum:
\begin{equation} \label{eq:Q0}
  Q_0(p) = \frac{\xi_{\rm inj} E_{\rm SN} \mathcal{R}_{\rm SN}(R)}{4 \pi \Lambda c (m_p c)^4}  
  		{\left( \frac{p}{m_p c} \right)}^{-\gamma}   \,.
\end{equation}
Here $E_{\rm SN} = 10^{51}$ erg is the total kinetic energy released by a single supernova explosion, $\xi_{\rm inj}$ is the fraction of such energy channeled into CRs, $\mathcal{R}_{\rm SN}(R)$ is the SN explosion rate per unit area and the normalization constant is $\Lambda= \int_{p_{\min}}^{p_{\max}} dy y^{2-\gamma} \left[ (y^2+1)^{1/2}-1 \right] $.

A standard technique to solve the transport equation (\ref{eq:transport}) is to integrate between $0^-$ and $0^+$ and between   $0^+$ and $z$ with the boundary conditions $f(0,p)=f_0(p)$ and $f(H,z)=0$ \cite[see, e.g.,][]{2012PhRvL.109f1101B,Recchia16}. One gets the following result for $f(z,p)$:
\begin{equation} \label{eq:f(z,p)}
  f(z,p) = f_0(p) \, \frac{1-e^{-\xi(z,p)}}{1-e^{-\xi(0,p)}}  \,,
\end{equation}
where $\xi(z,p) = \int_{z}^H v_A / D(z',p) dz'$ and the distribution function in the disk is:
\begin{equation} \label{eq:f0}
  f_0(p) = \int_{p}^{p_{\max}} \frac{dp'}{p'} \, \frac{3 Q_0(p')}{2 v_A} 
  	       \exp \left[ - \int_{p}^{p'} \left(  \frac{d \hat p}{\hat p} \frac{3}{e^{\xi(0,\hat p)} -1}  \right) \right]  \,.
\end{equation}

When magnetic perturbations are weak ($\delta B \ll B_0$), one can use quasi-linear theory to determine the diffusion coefficient, which is usually written as
\begin{equation} \label{eq:D}
  D(z,p) = D_B  \left[ \frac{1}{\mathcal{F}(k)} \right]_{k= 1/r_L}  \,,
\end{equation}
where $D_B=r_L v/3$ is the Bohm diffusion coefficient, with $r_L$ the Larmor radius and $v$ the particle's speed. $\mathcal{F}(k)$ is the normalized energy density per unit logarithmic wavenumber $k$, calculated at the resonant wavenumber $k= 1/r_L$. The local value of $\mathcal{F}$ is determined by the balance between the CR-driven growth of Alfv\'en waves and their damping. The growth rate due to the CR-streaming instability is \citep{Skilling75}:
\begin{equation} \label{eq:Gamma_cr}
  \Gamma_{\rm cr} = \frac{16 \pi^2}{3} \frac{v_A}{\mathcal{F}(k) B_0^2} 
  				\left[ p^4 v(p) \frac{\partial f}{\partial z} \right]_{p= e B_0/kc}  \,,
\end{equation}
while the dominant damping process in a region where the background gas is totally ionized is the non-linear Landau Damping (NLLD) that occurs at a rate \citep{2003A&A...403....1P}:
\begin{equation} \label{eq:Gamma_damp}
  \Gamma_{\rm nlld} = (2 c_k)^{-3/2} \, k v_A  \, {\mathcal{F}(k)}^{1/2} \,,
\end{equation}
with $c_k= 3.6$.
Since damping is much faster than wave advection at the Alfv\'en speed, a good approximation for the wave distribution can be obtained by equating  $\Gamma_{\rm cr}$ with $\Gamma_{\rm nlld}$, which returns the following implicit form for the wave spectrum:
\begin{equation} \label{eq:F}
  \mathcal{F}(k) = (2 c_k)^{3} \left[ \frac{16 \pi^2 p^4}{B_0^2} \, D \frac{\partial f}{\partial z} \right]^{2} \,.
\end{equation}
Inserting the diffusive flux, $D \partial f/\partial z$ (from equation (\ref{eq:f(z,p)})) into equation (\ref{eq:F}), we can derive the diffusion coefficient using equation (\ref{eq:D}):
\begin{equation} \label{eq:D(z,p)}
  D(z,p) = D_{\rm H}(p) + 2 v_A \left(H-z \right)  \,,
\end{equation}
where $D_{\rm H}(p)$ is the diffusion coefficient at $z=H$ and reads:
\begin{equation} \label{eq:DH}
  D_{\rm H}(p) =  \frac{D_B}{(2 c_k)^3} 
  		\left[  \frac{B_0^2}{16 \pi^2 p^4} \frac{1-e^{-\xi(0,p)}}{v_A f_0(p)} \right]^2 \,.
\end{equation}
It follows that the diffusion coefficient is maximum at $z=0$ and decreases linearly with $z$.

The exact solutions, equations (\ref{eq:f(z,p)})-(\ref{eq:f0}) and (\ref{eq:D(z,p)})-(\ref{eq:DH}), can be written in an explicit form in the two opposite limits of diffusion-dominated and advection-dominated transport.
In particular, it is straightforward to verify that in the diffusion dominated case (i.e. when $v_A H \ll D_{\rm H}$) the leaky box solution is recovered. In fact in this limit $e^{-\xi(0,p)} \approx 1-v_A H/D_H$ and $D$ becomes constant in $z$, namely
\begin{equation} \label{eq:D_no_vA}
  D(z,p) \rightarrow  D_{\rm H}(p) \rightarrow D_B^{\frac{1}{3}}  \frac{1}{2 c_k}
			\left[  \frac{B_0^2 H}{16 \pi^2 p^4  f_0(p)} \right]^{\frac{2}{3}}   \,.
\end{equation}
In the same limit equation (\ref{eq:f(z,p)}) reduces to the well known $f(z,p) = f_0(p) \left( 1- z/H \right)$, where $f_0(p) = Q_0 H/(2 D_H)$.  Replacing this last expression for $f_0$ into equation (\ref{eq:D_no_vA}) we obtain explicit expressions for both $D_{\rm H}$ and $f_0$, which read

\begin{equation} \label{eq:DH_diff}
  D_{\rm H}(p) =  D_B  \frac{1}{(2 c_k)^3}  \left[ \frac{2 B_0^2}{16 \pi^2 p^4 Q_0} \right]^2 
  		       \propto p^{2\gamma-7}
\end{equation}
and
\begin{equation} \label{eq:f0_diff}
  f_0(p) =  \frac{Q_0}{2} \frac{H}{D_H}
            =  \frac{3 c_k^3}{r_L v}  \left( \frac{16 \pi^2 p^4}{B_0^2} \right)^2  H Q_0(p)^3  
            \propto p^{7-3\gamma}\,,
\end{equation}
respectively. In the opposite limit, when $v_A H \gg D_{\rm H}$, we have that $e^{-\xi} \approx 0$, hence $f(z,p) \rightarrow f_0(p)$ and from equation~(\ref{eq:f0}) we recover:
\begin{equation} \label{eq:f0_adv}
  f_0(p) = \frac{3 Q_0(p)}{2 v_A \, \gamma} \,.
\end{equation}
On the other hand, we see from equation~(\ref{eq:D(z,p)}) that the diffusion coefficient in the disk behaves like a constant in momentum, namely $D(z=0,p) \rightarrow 2 v_A H$. This happen because for small $p$, $\mathcal{F}\rightarrow D_B/(2v_A H)$, hence $D=D_B/\mathcal{F} \rightarrow 2 v_A H$. Clearly this dependence is restricted to the momenta for which diffusion becomes comparable to advection, typically below $\sim 10$ GeV/c (see below). We refer to this regime as advection dominated regime, although particles never reach a fully advection dominated transport because diffusion and advection time are of the same order.

We are interested in describing the dependence of the CR spectrum on the Galactocentric distance, which enters the calculation only through the injection term and the magnetic field strength. Comparing equations (\ref{eq:f0_diff}) and (\ref{eq:f0_adv}) we see that the CR density has the following scalings with quantities depending on $R$:
\begin{eqnarray}
& f_0(p) \propto (Q_0/B_0)^3 \hspace{1.2 cm} \textnormal{(diffusive regime)} \nonumber \\
& f_0(p) \propto Q_0/B_0       \hspace{1.5 cm} \textnormal{(advective regime)}.
\end{eqnarray}

In a more general case, equations (\ref{eq:f0}), (\ref{eq:D(z,p)}) and (\ref{eq:DH}) can be solved iteratively. We start by choosing a guess function for $D_H(p)$ (for instance the expression, equation (\ref{eq:DH_diff}), obtained without advection) and then we iterate until convergence is reached, a procedure which usually requires only few iterations. Notice that the general case of a transport equation (\ref{eq:transport}) where the advection speed may depend on the $z$-coordinate has been recently discussed by \citep{Recchia16} and used to describe CR-induced Galactic winds. For the sake of simplicity, and to retain the least number of parameters, here we assume that the advection velocity is simply the Alfv\'en speed of self-generated waves and we assume that $v_{A}$ is independent of $z$.

\section{Results} 
  \label{sec:results}
\subsection{Fitting the local CR spectrum} \label{sec:localCR}

The rate of injection of CRs per unit surface can be calibrated to reproduce the energy density and spectrum of CRs as observed at the Earth. In all our calculations, following most of current literature, we choose a size of the halo $H=4$ kpc, while the ion density in the halo is fixed as $n_i = 0.02$ cm$^{-3}$, a value consistent with the density of the warm ionized gas component \cite[see, e.g,][]{Ferriere01}. The magnetic field $B_{0}$ at the location of the Sun is assumed to be $B_\odot=1 \mu$G (since we are only describing the propagation in the $z$ direction, this should be considered as the component of the field perpendicular to the disc). Notice that, given $B_0$ and $n_i$, the value of the Alfv\'en speed is also fixed, and this is very important in that it also fixes the momentum where the transition from advection propagation to diffusion dominated propagation takes place for a given injection spectrum and product of injection efficiency times the local SN explosion rate, $\xi_{\rm inj} \times \mathcal{R}_{\rm SN}(R_\odot)$. Following \cite[]{2012PhRvL.109f1101B,2013JCAP...07..001A,Aloisio15} we adopted a slope at injection $\gamma= 4.2$. Then, by requiring that the local CR density at $\sim 10-50$ GeV is equal to the observed one, we get $\xi_{\rm inj}/0.1 \times\mathcal{R}_{\rm SN}/(1/30 \, \rm yr) = 0.29$.

It is worth stressing that the CR spectrum in the energy region $\gtrsim 100$ GeV, may be heavily affected by either pre-existing turbulence \cite[]{2012PhRvL.109f1101B,2013JCAP...07..001A,Aloisio15} or a $z$-dependent diffusion coefficient \citep{Tomassetti}. Both possibilities have been proposed to explain the spectral hardening observed in both the protons' and helium spectrum at rigidities above $\sim 200$ GV. For this reason, a model including only self-generated diffusion can be considered as reliable only below $\sim 50$ GeV. In the following, we limit our attention to CRs that are responsible for the production of $\gamma$-rays of energy $\sim 2$ GeV, as observed by \fermilat, namely protons with energy of order $\sim 20$ GeV. This threshold is sufficiently low that the slope derived by ignoring the high energy spectral hardening can be considered reliable.


The injection parameters (efficiency and spectrum) found by fitting the CR density and spectrum at the Sun's location are assumed to be the same for the whole Galaxy. As discussed in \S \ref{sec:nonlocalCR}, the rate of injection of CRs per unit surface is then proportional to the density of SNRs as inferred from observations.

\subsection{CR spectrum in the Galactic disk} \label{sec:nonlocalCR}

The SNR distribution is usually inferred based on two possible tracers: radio SNRs and pulsars. Here we adopt the distribution of SNRs recently obtained by \cite{Green15} from the analysis of bright radio SNRs. He adopted a cylindrical model for the Galactic surface density of SNRs as a function of the Galactocentric radius, in the form:
\begin{equation} \label{eq:sources}
  f_{\rm SNR} \propto \left( \frac{R}{R_\odot} \right)^{\alpha} \, \exp\left( -\beta \frac{R-R_\odot}{R_\odot} \right) \,,
\end{equation}
where the position of the Sun is assumed to be at $R_\odot= 8.5$ kpc. For the best fit \cite{Green15} obtained $\alpha = 1.09$ and $\beta= 3.87$, so that the distribution is peaked at $R=2.4$ kpc. However, as noted by \cite{Green15}, it is worth keeping in mind that the best-fitting model is not very well defined, as there is some level of degeneracy between the parameters $\alpha$ and $\beta$.

\cite{CB98} also adopted a fitting function as in equation (\ref{eq:sources}) but obtained their best fit for $\alpha = 2.0$ and $\beta= 3.53$, resulting in a distribution peaked at $R=4.8$ kpc and broader for larger values of $R$ with respect to the one of \cite{Green15}. \cite{CB98} estimated the source distances using the so called `{\it $\Sigma$--D}' relation, that is well known to be affected by large uncertainties. Moreover \cite{Green15} argued that the {\it $\Sigma$--D} used by \cite{CB98} appears to have been derived incorrectly.

An important caveat worth keeping in mind is that the SNR distribution derived in the literature is poorly constrained for large galactocentric radii. For instance \cite{Green15} used a sample of 69 bright SNRs but only two of them are located at galactic latitude $l>160^{\circ}$. Similarly, \cite{CB98} used a larger sample with 198 SNRs, but only 7 of them are located at $R>13$ kpc and there are no sources beyond 16 kpc.

The distribution of pulsars is also expected to trace that of SNRs after taking into account the effect of birth kick velocity, that can reach $\sim 500$ km/s. These corrections are all but trivial, \citep[see, e.g.][]{Faucher06}, hence in what follows we adopt the spatial distribution as inferred by \cite{Green15}.

\begin{figure}
\begin{center}
\includegraphics[width=0.47\textwidth]{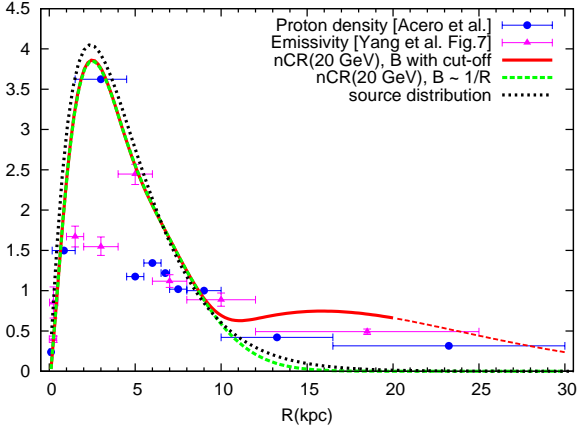}
\end{center}
\caption{CR density at $E>20$ GeV \citep{Acero16} and emissivity per H atom \citep{Yang16} as a function of the Galactocentric distance, as labelled. Our predicted CR density at $E>20$ GeV is shown as a dashed line. The case of exponentially suppressed magnetic field is shown as a solid line. The dotten line shows the distribution of sources \citep[]{Green15}.}
\label{fig:density}
\end{figure}

\begin{figure}
\begin{center}
\includegraphics[width=0.47\textwidth]{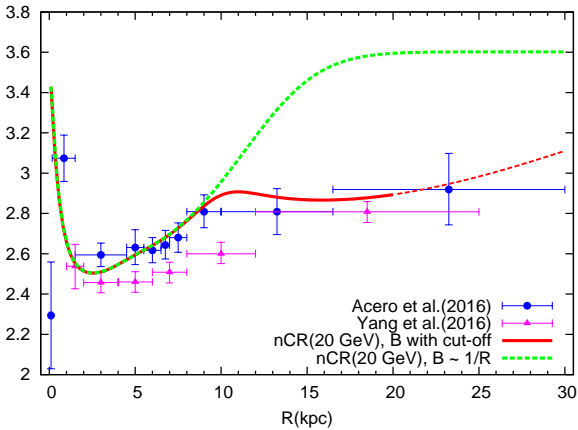}
\end{center}
\caption{Radial dependence of the power-law index  of the proton spectrum as inferred by \citep[][filled circle]{Acero16} and  \citep[][filled triangle]{Yang16}. Our predicted slope for the basic model is shown as a dashed line, while the solid line illustrates the results for the exponentially suppressed magnetic field.}
\label{fig:slope}
\end{figure}

One last ingredient needed for our calculation is the magnetic field strength, $B_0(R)$, as a function of galactocentric distance $R$. While there is a general consensus that the magnetic field in the Galactic disk is roughly constant in the inner region, in particular in the so-called ``molecular ring'', between 3 and 5 kpc \citep{Jansson12,stanev97}, much less is known about what the trend is in the very inner region around the Galactic center, and in the outer region, at $R>5$ kpc. Following the prescription of \cite{Jansson12}  \cite[see also][]{stanev97}, we assume the following radial dependence:
\begin{eqnarray} \label{eq:B0in}
B_0(R <5 \,{\rm kpc})  =  B_\odot R_\odot/{\rm 5 \, kpc}   \nonumber \\ 
B_0(R >5 \,{\rm kpc})  =  B_\odot R_\odot/R  \,,
\end{eqnarray}
where the normalization is fixed at the Sun's position, that is $B_\odot = 1\mu$G. Using this prescription we calculate the CR spectrum as a function of the Galactocentric distance, as discussed in \S \ref{sec:CR}. In Fig.~\ref{fig:density} we plot the density of CRs with energy $\gtrsim 20$ GeV (dashed line) and compare it with the same quantity as derived from \fermilat data. Our results are in remarkably good agreement with data, at least out to a distance of $\sim 10$ kpc. At larger distances, our predicted CR density drops faster than the one inferred from data, thereby flagging again the well known CR gradient problem. In fact, the non-linear theory of CR propagation, in its most basic form (dashed line) makes the problem even more severe: where there are more sources, the diffusion coefficient is reduced and CRs are trapped more easily, but where the density of sources is smaller the corresponding diffusion coefficient is larger and the CR density drops. A similar situation can be seen in the trend of the spectral slope as a function of $R$, plotted in Figure~(\ref{fig:slope}). The dashed line reproduces well the slope inferred from \fermilat data out to a distance of $\sim 10$ kpc, but not in the outer regions where the predicted spectrum is steeper than observed. It is important to understand the physical motivation for such a trend: at intermediate values of $R$, where there is a peak in the source density, the diffusion coefficient is smaller and the momenta for which advection dominates upon diffusion is higher. This implies that the equilibrium CR spectrum is closer to the injection spectrum, $Q(p)$ (harder spectrum). On the other hand, for very small and for large values of $R$, the smaller source density implies a larger diffusion coefficient and a correspondingly lower momentum where advection dominates upon diffusion. As a consequence the spectrum is steeper, namely closer to $Q(p)/D(p)$. In fact, at distances $R\gtrsim 15$ kpc, the spectrum reaches the full diffusive regime, hence $f_0 \sim p^{7-3\gamma} = p^{-5.6}$, meaning that the slope in Figure ~(\ref{fig:slope}) is 3.6. As pointed out in \S \ref{sec:CR}, the non-linear propagation is quite sensitive to the dependence of the magnetic field on $R$.

Both the distribution of sources and the magnetic field strength in the outer regions of the Galaxy are poorly known. Hence, we decided to explore the possibility that the strength of the magnetic field may drop faster than $1/R$ at large galactocentric distances. As a working hypothesis we assumed the following form for the dependence of $B_{0}$ on R, at $R\gtrsim 10$ kpc:
\begin{equation} \label{eq:B0out}
B_0(R> 10 \,{\rm kpc})  =  \frac{B_\odot R_\odot}{R} \, \exp\left[- \frac{R-10 \, {\rm kpc}}{d} \right]  
\end{equation}
where the scale length, $d$, is left as a free parameter. We found that using $d= 3.1$ kpc, both the resulting CR density and spectral slope describe very well the \fermilat data in the outer Galaxy. The results of our calculations for this case are shown in Figs. \ref{fig:density} and \ref{fig:slope} with solid lines.

\begin{figure}
\begin{center}
\includegraphics[width=0.47\textwidth]{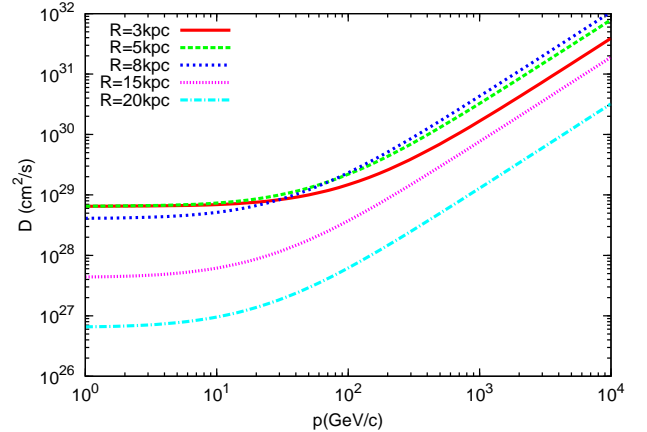}
\end{center}
\caption{Diffusion coefficient $D(z=0,p)$ as a function of momentum in GeV$/c$ for different Galactocentric distances as labelled.}
\label{fig:D}
\end{figure}

The diffusion coefficient resulting from the non-linear CR transport in the Galaxy, calculated as in \S \ref{sec:CR}, is illustrated in Fig.~\ref{fig:D}, for different galactocentric distances. It is interesting to notice that at all values of $R$ (and especially at the Sun's position) D(p) is almost momentum independent at $p\lesssim 10$ GeV/c. This reflects the fact that at those energies the transport is equally contributed by both advection and diffusion, as discussed above. This trend, that comes out as a natural consequence of the calculations, is remarkably similar to the one that in numerical approaches to CR transport is imposed by hand in order to fit observations.

Contrary to a naive expectation, in the case in which $B_{0}(R)$ drops exponentially, the diffusion coefficient becomes smaller in the external Galaxy than in the inner part, in spite of the smaller number of sources in the outer Galaxy. This counterintuitive result is due to the fact that $D_{\rm H}(p) \propto B_0^4/Q_0^2$ (see equation~\ref{eq:DH_diff}) and that both $B_0$ and $Q_0$ are assumed to drop exponentially at large $R$. Clearly, this result loses validity when $\delta B/B_0$ approaches unity and the amplification enters the non linear regime. Using equation (\ref{eq:D(z,p)}), such condition in the disk can be written as $\mathcal{F}(z=0,k) \approx D_B/(2v_A H)  \gtrsim 1$ which, for 1 GeV particles occurs for $R \gtrsim 28$ kpc (red-dashed line in Figures (\ref{fig:density}) and (\ref{fig:slope})). In any case, the density of CRs at large galactocentric distances drops down, as visible in Figure (\ref{fig:density}).

\section{Conclusions} \label{sec:conc}

The CR density recently inferred from \fermilat observations of the diffuse Galactic $\gamma$-ray emission, as carried out during the last seven years, appears to be all but constant with galactocentric distance $R$ \cite[]{Acero16,Yang16}. In the inner $\sim 5$ kpc from the Galactic center, such density shows a pronounced peak around $3-4$ kpc, while it drops with $R$ for $R\gtrsim 5$ kpc, but much slower than what one would expect based on the distribution of SNRs, as possible sources of Galactic CRs. Moreover, the inferred slope of the CR spectrum shows a gradual steepening in the outer regions of the Galaxy. This puzzling CR gradient is hard to accommodate in the standard picture of CR transport.  

Here we showed that both the gradient and the spectral shape can be explained in a simple model of non-linear CR transport: CRs excite waves through streaming instability in the ionized Galactic halo and are advected with such Alfv\'en waves. In this model, the diffusion coefficient is smaller where the source density is larger and this phenomenon enhances the CR density in the inner Galaxy. In the outer Galaxy, the data can be well explained only by assuming that the background magnetic field drops exponentially at $R\gtrsim 10$ kpc, with a suppression scale of $\sim 3$ kpc. This scenario also fits well the spectral slope of the CR spectrum as a function of $R$, as a result of the fact that at different $R$ the spectrum at a given energy ($\sim 20$ GeV) may dominated by advection (harder spectrum) or diffusion (softer spectrum). A simple prediction of our calculations is that the spectral hardening should disappear at higher energies, where transport is diffusion dominated at all galactocentric distances.

\section*{Acknowledgments}
The authors acknowledge useful conversations with R. Aloisio and C. Evoli.




\bibliographystyle{mnras}
\bibliography{biblio} 

%
%



\appendix




\bsp	
\label{lastpage}
\end{document}